\documentclass[aps,prd]{revtex4}
\usepackage{subfigure}
\usepackage{amssymb,amsmath,amsfonts,comment,float,latexsym,graphicx,epstopdf}
\usepackage{color}
\usepackage{fancyvrb}
\usepackage{chngcntr}

\newcommand{\be}{\begin{equation}}
\newcommand{\ee}{\end{equation}}
\newcommand{\ba}{\begin{eqnarray}}
\newcommand{\ea}{\end{eqnarray}}

\newcommand{\grts}{\raise.3ex\hbox{$>$\kern-.75em\lower1ex\hbox{$\sim$}}}
\newcommand{\lets}{\raise.3ex\hbox{$<$\kern-.75em\lower1ex\hbox{$\sim$}}}

\begin{document}
\title{The Electron/Muon Specific Two Higgs Doublet Model at $e^+e^-$ Colliders}

\author{Aria R. Johansen}\email[]{arjohansen@email.wm.edu}
\author{Marc Sher}\email[]{mtsher@wm.edu}

\affiliation{High Energy Theory Group, College of William and Mary, Williamsburg, Virginia 23187, U.S.A.}
\date{\today}

\begin{abstract}
Recently, Kajiyama, Okada and Yagyu (KOY) proposed an electron/muon specific two Higgs doublet model.  In this model, an $S_3$ symmetry suppresses flavor changing neutral currents instead of a $Z_2$ symmetry.   In the ``Type I" version of the model, the heavy Higgs bosons have a greatly enhanced coupling to electrons and muons.    KOY studied the phenomenology of the heavy Higgs bosons at the LHC.    In this paper, the phenomenology at electron-positron colliders is studied.  For the heavy Higgs mass range between $150$ and $210$ GeV, bounds from LEP-200 are stronger than those from the LHC.   The model allows for the interesting possibility that muon pair production at the ILC can be mediated by s-channel Higgs exchange.  This requires an energy scan.  The scanning rate and necessary resolution are discussed.
\end{abstract}

\maketitle

\section{Introduction}

One of the most widely studied extensions of the Standard Model is the Two Higgs Doublet Model (2HDM) (see Ref. \cite{Branco:2011iw} for a review).    A feature of 2HDMs is that they will generally have tree level flavor-changing neutral currents (FCNC).   Conventionally, a $Z_2$ symmetry is imposed which forces all of the fermions of a given charge to couple to a single Higgs multiplet.  This coupling of a fermion to a single Higgs multiplet can happen in one of several ways.    For the Type-I model, all fermions couple to a single multiplet, usually denoted $\Phi_2$.  The Type-II model, alternatively couples the charge $Q=-1/3$ quarks and leptons to the $\Phi_1$ multiplet and the $Q=2/3$ quarks to the $\Phi_2$ multiplet. Two more types, the Lepton-Specific and Flipped models, reverse the couplings of the leptons.

Recently, Kajiyama, Okada and Yagyu\cite{Kajiyama:2013sza} (KOY) proposed another version of the 2HDM.   In this version, an  $S_3$ symmetry suppresses the FCNC, instead of the more conventional $Z_2$ symmetry.       $S_3$  was chosen as it is the simplest non-Abelian discrete symmetry which constrains the Yukawa couplings.   One version of the model has the heavy Higgs bosons primarily decaying into electrons and muons, leading to a radically different phenomenology than conventional 2HDMs.   In the KOY paper, the authors found that with current LHC data,  the heavy Higgs bosons can be stringently bounded if their masses are less than $150$ GeV.

The LHC is not the only way to find constraints on this model.  This paper explores how electron-positron colliders can further constrain the model through s-channel Higgs exchange in muon pair production.  With a heavy Higgs that couples relatively strongly to electron and muon pairs, the measurement of the muon pair production cross section can further constrain the model's parameters. It will be shown that bounds from LEP-200 can be stronger than those from the LHC.  The sensitivity of the International Linear Collider(ILC) will also be discussed.
 
The relevant parts of the KOY model are reviewed in Section II.  Section III will be a discussion of the limits obtainable from LEP-200 data, and a comparison with the LHC reach obtainable in the KOY model. A discussion of the potential reach of the ILC, and the requirements for an energy scan to detect the s-channel resonance is in Section IV.  Section V contains the conclusions. 
  
 \section{The Model}
 
The model proposed by KOY assigns the first and second family of lepton fields (both the left-handed isodoublets and the right-handed isosinglets) to the doublet of $S_3$.   The third family of leptons, all the quark fields and the Higgs doublets are $S_3$ singlets.   $S_3$ contains two types of singlet irreducible representations, the $1$ and the $1^\prime$.    One of the Higgs doublets and the left-handed quark and tau doublets are assigned to the $1$ and the right-handed up quarks and the other Higgs doublet are assigned to the $1^\prime$.  There are four different versions depending on whether the right-handed down quarks and right-handed tau are assigned to the $1$ or $1^\prime$.   The version analogous to the Type-I 2HDM happens after assigning one of the Higgs doublets and the left-handed quark and tau doublets to the $1$, while the right-handed up quarks and other Higgs doublets are assigned to the $1^{\prime}$.    KOY note that the other assignments do not lead to phenomenology substantially different from the conventional Type-II, Lepton-Specific and Flipped models.

Limiting the focus of this paper to the Type-I case of the KOY model, and to only neutral scalar couplings to electrons and muons, the Yukawa Lagrangian contains (see Ref. \cite{Kajiyama:2013sza}  for details)
\ba
\label{eq:Lagrang}
{\cal L}_Y \supset \frac{m_\mu}{v} &\Big( & \frac{1}{2}(\tan\beta + \cot\beta)c_{\beta-\alpha}\bar{e}eh - [s_{\beta-\alpha}-\frac{1}{2}(\tan\beta - \cot\beta)c_{\beta-\alpha}]\bar{\mu}\mu h \cr\cr
&-&  \frac{1}{2}(\tan\beta + \cot\beta)s_{\beta-\alpha}\bar{e}eH - [c_{\beta-\alpha}+\frac{1}{2}(\tan\beta - \cot\beta)s_{\beta-\alpha}]\bar{\mu}\mu H \cr\cr
&+& \frac{i}{2}(\tan\beta+\cot\beta)\bar{e}\gamma_5 e A + \frac{i}{2}(\tan\beta-\cot\beta)\bar{\mu}\gamma_5\mu A \Big)
\ea

Here, the electron mass is neglected, $s_{\beta-\alpha} (c_{\beta-\alpha})$ are $\sin(\beta - \alpha)$ ($\cos(\beta-\alpha)$) where $\tan\beta$ is the ratio of the two vacuum expectation values and $\alpha$ is the angle diagonalizing the neutral scalar mass matrix.  $h$ is the light Standard Model Higgs, with a mass of $126$ GeV and $H$ and $A$ are the heavy neutral scalar and pseudoscalar.    Since the $HZZ$ and $hZZ$ couplings are given, relative to the Standard Model couplings, by $\cos(\beta-\alpha)$ and $\sin(\beta-\alpha)$ respectively, and since the latter is not too far off the Standard Model value, KOY chose $\sin(\beta-\alpha)=1$, in which case the $126$ GeV state has the same coupling to vectors and fermions as the Standard Model.  This choice will be adopted initially here, but a discussion of relaxing it appears later in this paper.    Note that the model is somewhat fine-tuned, in that the electron and muon masses come from the sum and difference of two similar terms.

In all conventional 2HDMs, the muon and electron Yukawa couplings are suppressed by factors of $m_\mu/v$ and $m_e/v$, respectively and can generally be neglected.   In addition, in Type-I models, the couplings of the $H$ and $A$ to the other fermions are suppressed by a factor of $1/\tan\beta$.    With the $S_3$ model, the couplings of the $H$ and $A$ to the electron and muon are  suppressed by a factor of $m_\mu/v$,   the couplings to all other fermions are also suppressed by $1/\tan\beta$, but the couplings to the electron and muon are now enhanced by a factor of $\tan\beta$.    This then causes, for large $\tan\beta$, the coupling of the electrons and muons to be dominant.  KOY show this, and plot the branching ratios of $H$ and $A$ as a function of $\tan\beta$, showing that decays into muons and electrons are dominant for $\tan\beta$ larger than 10.

KOY studied the phenomenology of the model at the LHC.    They considered gluon fusion into an $H$ or an $A$, and the subsequent decay into muons. KOY also studied ATLAS searches for Higgs bosons from the dimuon decay signature.   For small $\tan\beta$, the branching ratio into muons is small, and for large $\tan\beta$, the production cross section from gluon fusion is suppressed.   They thus rule out a range of $\tan\beta$ from approximately $3.0 - 25.0$ for $H$ and $A$ masses between $110$ and $150$ GeV.   In this analysis, they assume that the additional Higgs bosons are degenerate in mass, but the results depend only weakly on this assumption.

In addition,   they consider pair production $pp\rightarrow HA$ and $pp\rightarrow H^\pm H/H^\pm A$, with subsequent decay to four leptons (either four muons or three muons and a neutrino).   They conclude that the LHC has excluded all Higgs masses below $140$ GeV for all values of $\tan\beta \geq 3$.    These results are much more sensitive to the assumption that the $H$, $A$, $H^\pm$ masses are very similar, of course.

In all cases, no bounds were found for $H$ or $A$ masses greater than $150$ GeV.  They do show that these bounds will be substantially improved during the next couple of runs at the LHC, giving projections for $300$ and $3000$ inverse femtobarns.

This article considers constraints and projections from linear colliders, focusing on the specific process of muon pair production:  $e^+ e^- \rightarrow H/A \rightarrow \mu^+ \mu^-$.    In conventional 2HDMs, s-channel Higgs exchange is negligible due to the small electron coupling to the Higgs, and thus s-channel Higgs exchange can only be detected at a muon collider.   In the KOY model, however, the electron coupling is similar to the muon coupling, and thus s-channel Higgs exchange can be studied at an electron collider.

\section{Constraints from LEP-200}

The bounds obtained by KOY only restricted $\tan\beta$ for heavy Higgs masses below $150$ GeV.  The relatively large coupling of the $H$ and $A$ to electrons in the model allows for the bounds to be studied at an electron-positron collider. This will give bounds up to the kinematic limit of the collider.

In the $e^+ e^- \rightarrow H/A \rightarrow \mu^+ \mu^-$ process, resonances occur at $\sqrt{s} = m_H, m_A$.  For a heavy Higgs mass of $200$ GeV and $\tan\beta = 20$, the width of the Higgs in this model is approximately $1.0$ MeV, and scales as the mass and as $\tan^2\beta$.   For the regions of interest, this will always be substantially less than the beam spread, and will lead to a resonant enhancement in the signal's amplitude.   Detection of the process, as in detection of the Standard Model Higgs at a muon collider, requires an energy scan.

The LEP-200 accelerator measured the $e^+ e^- \rightarrow \mu^+ \mu^-$ process.   They did not do a comprehensive energy scan, as that was not the purpose of the experiment.  However, LEP-200 did run at twelve distinct energies of $\sqrt{s}$.   The combined LEP results for all four experiments are given in Figure 1 \cite{lep2results}.  One can see that there is a large gap in coverage between $136$ and $161$ GeV, with closer and closer separations as the kinematic limit of the collider is reached.  In Ref. \cite{lep2results}, the Standard Model expectations are also given for each of these energy values.   These expectations not only include tree-level $Z$ and $\gamma$ exchange, but also include radiative corrections.

\begin{figure}[h]\label{fig:Lep}
  \begin{center}
    \begin{tabular}{ |c | c || c | c | }
      \hline
      $\sqrt{s}$ (GeV) & Average Value (pb) & $\sqrt{s}$ (GeV) & Average Value (pb) \\ \hline \hline
      130 & $8.62\pm 0.68$ & 136 & $8.27\pm 0.67$  \\ \hline
      161 & $4.61\pm 0.36$ & 172 & $3.57\pm 0.32$  \\ \hline
      183 & $3.49\pm 0.15$ & 189 & $3.123\pm 0.076$ \\ \hline
      192 & $2.92\pm 0.18$ & 196 & $2.94\pm 0.11$ \\ \hline
      200 & $3.02\pm 0.11$ & 202 & $2.58\pm 0.14$ \\ \hline
      205 & $2.45\pm 0.10$ & 207 & $2.595\pm 0.088$ \\
      \hline
    \end{tabular}
  \end{center}
  \caption{Experimental results for muon pair production at LEP200}
\end{figure}

The process $e^+ e^- \rightarrow H/A \rightarrow \mu^+ \mu^-$ further constrains the KOY model.  The  interference between this process and tree-level Standard Model processes is proportional to the electron mass and is thus negligible.   Thus the cross-section is added to the Standard Model result to give the new cross section.    Compared with the experimental results, the bounds are obtained by requiring that the difference with experiment be less than $2$ sigma.

 For the $\sin (\beta-\alpha)=1$ limit, a plot showing the restricted regions of parameter space is shown in Figure 2.   The curve corresponds to the upper limit on $\tan\beta$ as a function of the $H$ or $A$  mass.   One can clearly see the resonances at the experimental $\sqrt{s}$ values.  If $m_H$ or $m_A$ are precisely on resonance, the signal is so large that the model is excluded for all $\tan\beta$.   Results apply for either $H$ and $A$, since the couplings are identical.  Due to the resonance structure, interference between $H$ and $A$ would be destructive, and will not occur unless the masses are very close to each other.  It is assumed here that this is not the case.
 
\begin{figure}[h]
  \begin{center}
    \includegraphics[scale=1.3]{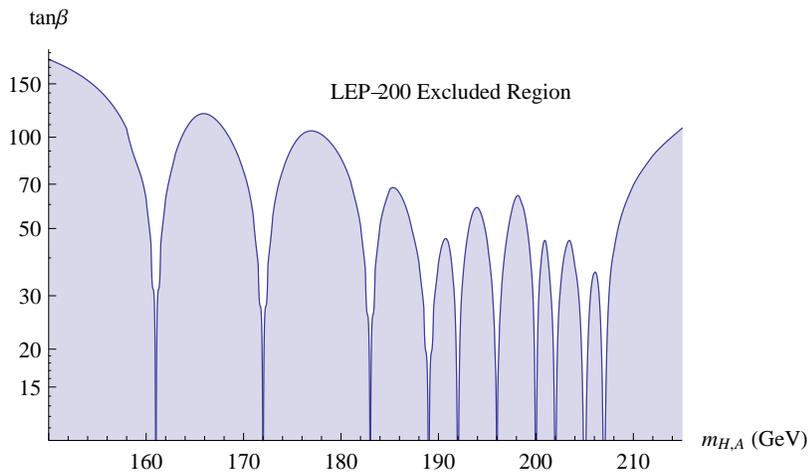}
  \end{center}
  \caption{The white region of parameter space is ruled out by LEP-200 results}
\end{figure}
\label{fig:sinBA1}
 
The bounds are quite weak for Higgs masses of $140-160$ GeV, due to the sparse coverage of LEP-200 in that energy region.  Above $160$ GeV, the bounds can be significant up to the maximum energy of the LEP-200 collider, $207$ GeV.  The masses below $140$ GeV were excluded by KOY.  These results put constraints on masses for the KOY model's $H$ and $A$ bosons for the masses between $150$ and $215$ GeV, for which there are no LHC bounds.

These calculations have been made assuming that $\sin(\beta-\alpha) =1$.  Supposing this is not the case, the coupling of the light Higgs, $h$ to electrons is no longer proportional only to the electron mass.   However, since the focus here is on masses above $160$ GeV, the propagator enhancement will be small, thus interference between the $h$ and $H/A$ exchange diagrams will be negligible.  The couplings of both the $H$ and the $A$ to electrons and muons depends on $\sin(\beta-\alpha)$, as can be seen in Equation~\ref{eq:Lagrang}. 

Current limits on $\sin(\beta-\alpha)$ from Type-I models \cite{Chen:2013rba, Dumont:2014wha,Chang:2013ona,Craig:2013hca} are rather weak, corresponding to $\sin(\beta-\alpha) > 0.85$.   These results are not based on the electron or muon couplings of the Higgs, and thus they would apply in this model as well.   In the KOY model, a nonzero $\cos(\beta-\alpha)$ will also lead to a different coupling of the light Higgs, $h$, to muons, leading to an enhanced $h \rightarrow \mu^+\mu^-$ rate.   The current bound from the LHC is less than $7.4$ times the Standard Model rate \cite{Khachatryan:2014jya}.   This means that 
\be
\sin(\beta-\alpha) - \frac{1}{2}(\tan\beta - \cot\beta)\cos(\beta-\alpha) < \sqrt{7.4}
\ee
For $\tan\beta = 14\ (16,24,37,54)$, this corresponds to $\sin(\beta-\alpha) > 0.85 (0.9,0.95,0.98,0.99)$.   For $\sin(\beta-\alpha)\neq 1$, $\tan\beta$ on the y-axis of Figure 2 is replaced  with $2\cos(\beta-\alpha) +\tan\beta\sin(\beta-\alpha)$, in accordance with Equation~\ref{eq:Lagrang}. This result is very similar to multiplying the scale on the y-axis by $\sin(\beta-\alpha)$, since the majority of the excluded region is for $\tan\beta>10$. Since $\sin(\beta-\alpha)>0.9$, the resulting figures for values of $\sin(\beta-\alpha)\neq 1$ are almost identical to the plot given in Figure 2, but with weaker bounds.

\section{Expectations from the ILC}

 In this section, the possibility of covering the entire allowed region of the $S_3$ 2HDM model up to the kinematic limit of the ILC is explored.

From Figure 2, it can be seen that if LEP-200 had done an energy scan over the entire region, the bounds on the model would have been stronger.   Such an energy scan may be possible at the ILC.  One of the main motivations for a muon collider is the s-channel production of Standard Model Higgs allowing at high precision studies\cite{Barger:1996pv}.  The resolution of the muon collider is expected to be small, $0.01\%$, thus making it is necessary to have information about the heavy Higgs mass prior to beginning the scan.  See Ref. \cite{Barger:1996pv} for a detailed analysis.  The resolution of the ILC is expected to be of the order of $0.1\%$.  This higher resolution reduces the signal, but allows for a more extensive scan.   

The analysis of Ref. \cite{Barger:1996pv} for the muon collider will be followed closely.    Using the ILC Technical Design Report, and combining the expected resolutions of the beams in quadrature \cite{ilctdr}, the center of mass energy ranges from $250$ GeV to $500$ GeV, causing the resolution to vary from  $0.11\%$ to $0.09\%$, respectively.  The value of $0.1\%$ is chosen.    This gives an RMS spread in the center of mass energies, $\sqrt{s}$.  This RMS spread is denoted by $\sigma_{\sqrt{s}}$, of
\be
\sigma_{\sqrt{s}} = R\sqrt{s}/\sqrt{2} = 140\ {\rm MeV} \left( \frac{\sqrt{s}}{200\ {\rm GeV}}\right)
\end{equation}
For the muon collider, $\sigma_{\sqrt{s}}$ was comparable to the Higgs width.   Here, however, it is much larger, and thus one can look at the effective signal cross section for $\sqrt{s} = m_H$.  Note that the results are unchanged if $H$ is replaced by $A$.  Following \cite{Barger:1996pv}, 
\begin{equation}
\sigma_H = \frac{2\pi^2 \Gamma(H \rightarrow ee) BF (H \rightarrow \mu\mu)}{m_H^2} \times \frac{1}{\sigma_{\sqrt{s}}\sqrt{2\pi}}.
\end{equation}
The branching fraction for $H\rightarrow\mu\mu$, as given by KOY \cite{Kajiyama:2013sza}, is approximately $0.5$ for $\tan\beta\ge 20$, and falls roughly as $\tan^2\beta$ for smaller $\tan\beta$.  This holds for the entire relevant mass range.   Note that the signal falls as the RMS spread in $\sqrt{s}$ increases.    The partial width $\Gamma(H\rightarrow ee)$ depends on the $H\bar{e}e$ Yukawa coupling.  Choosing $\tan\beta = 20$, $\sin(\alpha-\beta)=1$ and $m_H = 200$ GeV, the effective signal cross section is $15$ picobarns.

For the luminosity,  the expected \cite{ilctdr} value for the $250$ GeV ILC is $L_{250}=0.75 \times 10^{34}\ {\rm cm}^{-1}{\rm sec}^{-1}$.  The luminosities for higher energies are expected to be up to $2.4$ times larger.  The length of time needed to produce a five $\sigma$ signal can be found from the signal's cross section and the background for muon pair production.

Using $\tan\beta \le 20$, $\sqrt{s}=m_H= 200$ GeV, and $\sin(\alpha-\beta)=1$, the time needed to produce a five $\sigma$ signal is
\be
t = (20\ {\rm seconds})\left(\frac{20}{\tan\beta}\right)^8
\ee
 The sensitivity to $\tan\beta$ arises as each vertex depends approximately linearly on $\tan\beta$.  This makes the cross section scale as $\tan^4\beta$, and the needed time scale as $\tan^8\beta$.  To achieve the sensitivity of $\tan\beta=5$ would take two weeks.

For the muon collider, the RMS spread is approximately 7 MeV.  This makes it necessary to have information about the Higgs mass (within a few GeV) before scanning the possible energy range.  With an RMS spread ranging from $100$ MeV to $350$ MeV over the range of $\sqrt{s}$, knowledge of the heavy Higgs mass within a few GeV would only require 5-10 different energies to scan the relevant range.  Scanning the entire mass range from $150-500$ GeV would require $1000$ different energies and would not be feasible, given the length of time needed to change the energy setting.  As noted by KOY, future runs at the LHC will be sensitive to much higher masses.  With this information, it is thinkable that one or more of the heavy Higgs bosons will be discovered.  Knowing the masses of the newly discovered heavy Higgs bosons will then allow the ILC to do precision measurements.

More work needs to be done to establish the sensitivity of an energy scan at the ILC.   Initial state radiation will distort the shape of the resonance.  It should be possible to scan without needing (in the first pass) a five sigma discovery.   The angular distribution of the muons in the signal will be different from those in the background.   For small values of $\tan\beta$, the muon pair signal is small; however the $\bar{b}b$ signature would be detectable near resonance.  Similarly, a Higgs heavier than $350$ GeV will produce a $\bar{t}t$ signature.  Should this model receive experimental support at the LHC, a detailed investigation of this process at the ILC will be important.

\section{Concluding Remarks}
One of the most recently developed 2HDMs is the model of the KOY collaboration in which an $S_3$ symmetry is used in place of the usual $Z_2$ symmetry.    The Type I version of the model is of interest because the couplings of the heavy Higgs bosons (the $H$, $A$ and $H^\pm$) to the electron and the muon are strongly enhanced relatively to usual 2HDMs. Decays to electrons and muon can then dominate.  KOY studied the phenomenology at the LHC.  

The s-channel heavy neutral Higgs mediated contribution to the $e^+e^-\rightarrow\mu^+\mu^-$ process at an electron collider allows for further constraints to be placed on the model.  In the past, resonant s-channel Higgs exchange has only been considered for muon colliders, but in this model it can occur at an electron collider.   Bounds from LEP-200 are stronger than those obtained from the LHC for the $150-210$ GeV mass range.    These bounds are very sensitive to the precise $\sqrt{s}$ at which LEP-200 was run.
 
At the ILC, the beam resolution is substantially larger than that of the muon collider, and this will allow for the possibility of scanning over a broad range of masses.  Due to the resonant structure, it is crucial for $\sqrt{s}$ to be close to the Higgs mass.  The potential reach of such a scan was discussed.  
 
 \vskip 1.0cm
{\parindent = 0pt {\bf Acknowledgments}}
 \vskip 0.5cm
The work of A.J. was supported by the NSF under Grant PHY-1068008.    Both authors thank Chris Carone, Josh Erlich, and Mike Kordosky for their helpful discussions.

\end{document}